\documentclass{pasj00}

\begin{document}
\SetRunningHead{S. Shelyag, D. Przybylski}{Spectro-polarimetric signatures of photospheric Alfv\'en waves}

\title{Centre-to-limb spectro-polarimetric diagnostics of simulated solar photospheric magneto-convection: signatures of photospheric Alfv\'en waves}

\author{S. \textsc{Shelyag}
and 
D. \textsc{Przybylski}}
\affil{Monash Centre for Astrophysics, School of Mathematical Sciences, Monash University, Clayton, Victoria, 3800, Australia}
\email{sergiy.shelyag@monash.edu}


%


\maketitle

\begin{abstract}
Using numerical simulations of the magnetised solar photosphere carried out with the radiative magneto-hydrodynamic code, MURaM, and detailed 
spectro-polarimetric diagnostics of the simulated photospheric 6302$~\mathrm{\AA}$ FeI line, spectro-polarimetric signatures of Alfv{\'e}n waves 
in magnetised intergranular lanes of the simulated solar photosphere were analysed at different positions at the solar disk. The torsional 
Alfv{\'e}n waves in the intergranular lanes are horizontal plasma motions, which do not have a thermal perturbation counterpart. We find 
signatures of Alfv{\'e}n waves as small-scale line profile Doppler shifts and Stokes-$V$ area asymmetry enhancements in the simulated off-disk 
centre observations. These photospheric features disappear when the simulated observations are degraded with a telescope PSF similar to the one of Hinode. 
We analyse the possibilities for direct observations and confirmation of Alfv{\'e}n wave presence in the solar photosphere.
\end{abstract}

\section{Introduction}
Recently, the topic of swirling, torsional motions in the solar atmosphere has attracted a lot of attention in solar physics community as a possible
solution to the chromospheric and coronal heating problem \citep{wedemeyer1}. Observationally, small-scale swirling motions were found in the 
solar photosphere as magnetic bright point rotation \citep{bonet2008}, in the chromosphere as chromospheric swirls \citep{wedemeyer2009}, 
and as photospheric counterparts of incompressible waves in the quiet solar chromosphere \citep{morton2013}. It should, however, be mentioned 
that there is a growing evidence for swirling motions in the upper layers of the solar atmosphere (tornado-like prominences) to be a result of 
oscillations or a projection effect \citep{panasenco2014}.

In the simulations, swirling motions in the photospheric magnetic field concentrations were first identified by \citet{voegler1}. \citet{shelyag2011} 
numerically analysed mechanisms of photospheric vorticity generation and found that the major contribution in the vertical component of the vorticity vector 
in the photosphere is linked to magnetic tension in low plasma-$\beta$ regions of intergranular magnetic flux concentrations. Further numerical study 
ultimately demonstrated that apparent swirling motions in the magnetised intergranular lanes of the solar photosphere are short-lived \citep{moll2012},
torsional oscillations, which propagate with Alfv{\'e}n speed along the magnetic field: they are Alfv{\'e}n waves \citep{shelyag2013}. These horizontal 
torsional motions in the strong, low plasma-$\beta$ magnetic field generate positive (directed outwards from the Sun) Poynting flux \citep{shelyag2012},
which is, indeed, large enough to supply the solar corona with energy.

As it was mentioned before, Alfv{\'e}n waves are incompressible motions, which propagate along the magnetic field with the local Alfv{\'e}n speed. The Alfv{\'e}n
speed in the photospheric intergranular magnetic flux concentrations, which have a magnetic field strength of the order of $1~\mathrm{kG}$ 
\citep{martinez1997, shelyag2007}, reaches tens of kilometers per second \citep{fujimura2009, shelyag2013}. Thus, a simple estimate suggests
that it takes tens of seconds for an Alfv{\'e}n wave to propagate through the photosphere. Due to incompressibility of Alfv{\'e}n waves, there is
no observable variation in either photospheric continuum intensity, or in thermal variations of the photospheric absorption line profile shape.
While absorption line profile broadening was suggested as a signature of torsional motions \citep{van2008, jess2009}, rapid propagation
of  Alfv{\'e}n waves through the photosphere may leave them undetected. Therefore, an interesting question lies in observational confirmation of the 
presence of Alfv{\'e}n waves in the magnetic solar photosphere.

In this paper, we provide a preliminary analysis of spectro-polarimetric signatures of torsional Alfv{\'e}n waves, which propagate in the magnetised 
photosphere. Using a photospheric "plage" model, generated by the radiative magneto-hydrodynamic code MURaM \citep{voegler1},
and spectro-polarimetric diagnostics of the simulated model at three different angles of inclination, representing three positions at the solar
disk, the presence of elongated structures in Stokes-$V$ asymmetries and in line Doppler shift maps is demonstrated, which can be related to the 
horizontal torsional motions of plasma in magnetised intergranular lanes, or photospheric Alfv{\'e}n waves.

The paper is organised as follows. In Section 2, we describe the simulations and radiative diagnostics tools we used to generate the spectro-polarimetric 
data for analysis. Analysis of the generated data and spectro-polarimetric signatures of Alfv{\'e}n waves are presented in Section 3. In Section 4
we demonstrate non-detectability of Alfv\'enic motions with current instruments for solar photospheric observations. Section 5 concludes our findings, 
where we also discuss possible ways for photospheric torsional Alfv{\'e}n wave observations.

\section{Codes and tools}
We use the MURaM code \citep{voegler1} to produce models of solar photospheric magneto-convection. The MURaM code is a magnetohydrodynamic (MHD) solver
on a Cartesian grid, which includes the effects of partial ionization through the equation of state and takes into account non-grey radiative energy
transport. A domain size of $12 \times 12~\mathrm{Mm}$ in the horizontal directions, and $2.56~\mathrm{Mm}$ in the vertical direction is used
in the simulations. A grid size of $480\times480\times256$ is used, which leads to spatial resolution of $25 \times 25~\mathrm{km}$ in the
horizontal, and $10~\mathrm{km}$ in vertical direction, respectively. A pre-tabulated equation of state for the solar chemical composition, based 
on FreeEOS \citep{irwin2012} is used. The side boundaries of the domain are periodic, the top boundary is closed, and the bottom boundary is open 
for in- and outflows.

We use a $200~\mathrm{G}$ unipolar initial magnetic field to simulate a solar plage region. Starting from a non-magnetic photospheric convection 
model, the magnetic field was introduced in the numerical domain and let to evolve for few granular lifetimes until reaching a statistically steady state
and standard value of solar radiative flux. During this initial stage of the simulation, the magnetic field was redistributed by convective flows into
the intergranular downflows and experienced convective collapse. This led to formation of magnetic flux concentrations with the strength of about
$1-1.5~\mathrm{kG}$. Then, we recorded a photospheric magneto-convection model,
which was used for spectro-polarimetric diagnostics with the absorption line profile synthesis code NICOLE (a detailed paper is forthcoming, but it 
is similar to the code used by \citet{socas2000, socas2011}). The code
was used in local thermodynamic equilibrium (LTE) mode. For the diagnostics, we computed the full Stokes vector for a single magnetically-sensitive 
absorption line of FeI at $6302.5~\mathrm{\AA}$.

The primary aim of this initial study is to find the observational signatures of torsional Alfv{\'e}n waves in the solar photosphere. These waves 
are incompressible, therefore they have no signatures in continuum intensity or in thermal variations of an absorption line profile shape. However,
it could be possible to detect torsional horizontal plasma motions in the nearly vertical photospheric magnetic field concentrations \citep{voegler1}
if the line-of-sight velocity had a horizontal velocity component. According to the equation for line-of-sight velocity 
\begin{equation}
v_{los}=v_z \cos\theta + v_h \sin\theta, 
\label{incline}
\end{equation}
where $v_z$ and $v_h$ are the perpendicular and parallel to the solar surface velocity components, respectively, and $\theta$ is the angle between
the line of sight and the perpendicular to the solar surface, this occurs at a non-zero angle from the solar disk centre due to the spherical shape of the 
Sun. The influence of the horizontal velocity component in $v_{los}$ grows with the $\theta$.

Thus, to identify observational signatures of Alfv{\'e}n-type motions and, in general, study the effects of different positions at the solar disk 
on spectro-polarimetric observations, we performed spectro-polarimetric diagnostics for the simulated photospheric model at three inclination angles 
$\theta=0^\mathrm{o}$, $30^\mathrm{o}$, and $60^\mathrm{o}$. Inclination of the line-of-sight was simulated by appropriate shifting of the
layers in the simulated photospheric model and recomputing the line-of-sight spatial grid step accordingly. The line-of-sight velocity was 
recalculated according to Eq.~\ref{incline}, and the line-of-sight and azimuthal magnetic field components were computed in a similar manner. 
It should also be noted that slanting of the model layers was carried out in the positive (right) direction, thus the equivalent line-of-sight inclination 
was in the negative (to the left from the solar disk centre) direction.

\section{Simulated spectro-polarimetry at different observational angles}

\begin{figure*}
\begin{center}
  \includegraphics{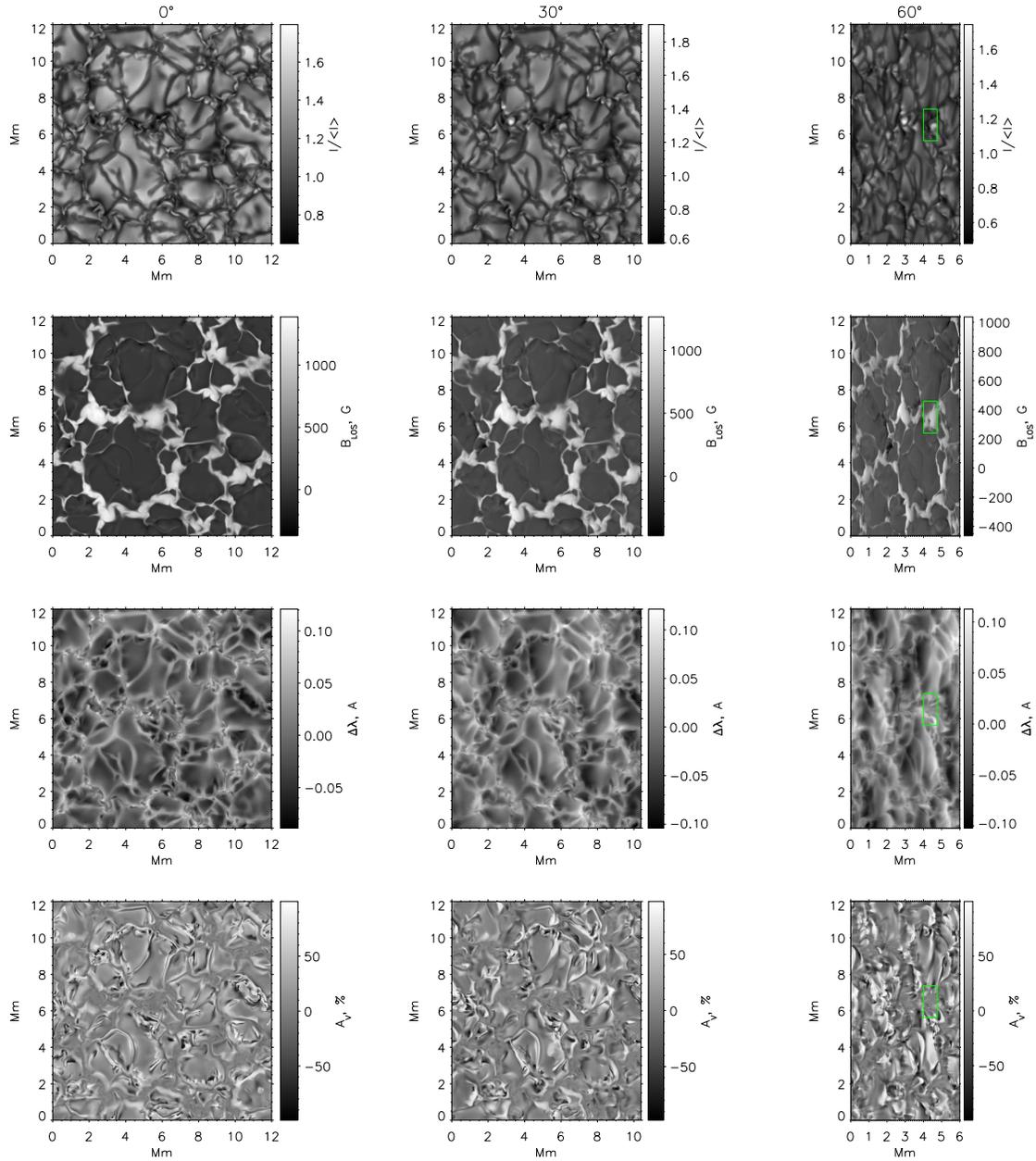} 
 \end{center}
\caption{An example of radiative parameters computed for the simulated magnetised photosphere. First row: normalized continuum intensity at 
6303\AA. Second row: line-of-sight magnetic field strength, measured at a constant geometric height, approximately the level of continuum formation. Third 
row: FeI 6302\AA~absorption line Doppler shift, computed by the centre of gravity method. Fourth row: FeI 6302\AA~Stokes-$V$ area asymmetry. The
maps are computed for three different inclinations, $0^o$ (first column), $30^o$ (second column) and $60^o$ (third column) at the solar disk. The region
studied in a greater detail is marked by a box in the right column.}
\label{fig1}
\end{figure*}

\begin{figure*}
\begin{center}
  \includegraphics{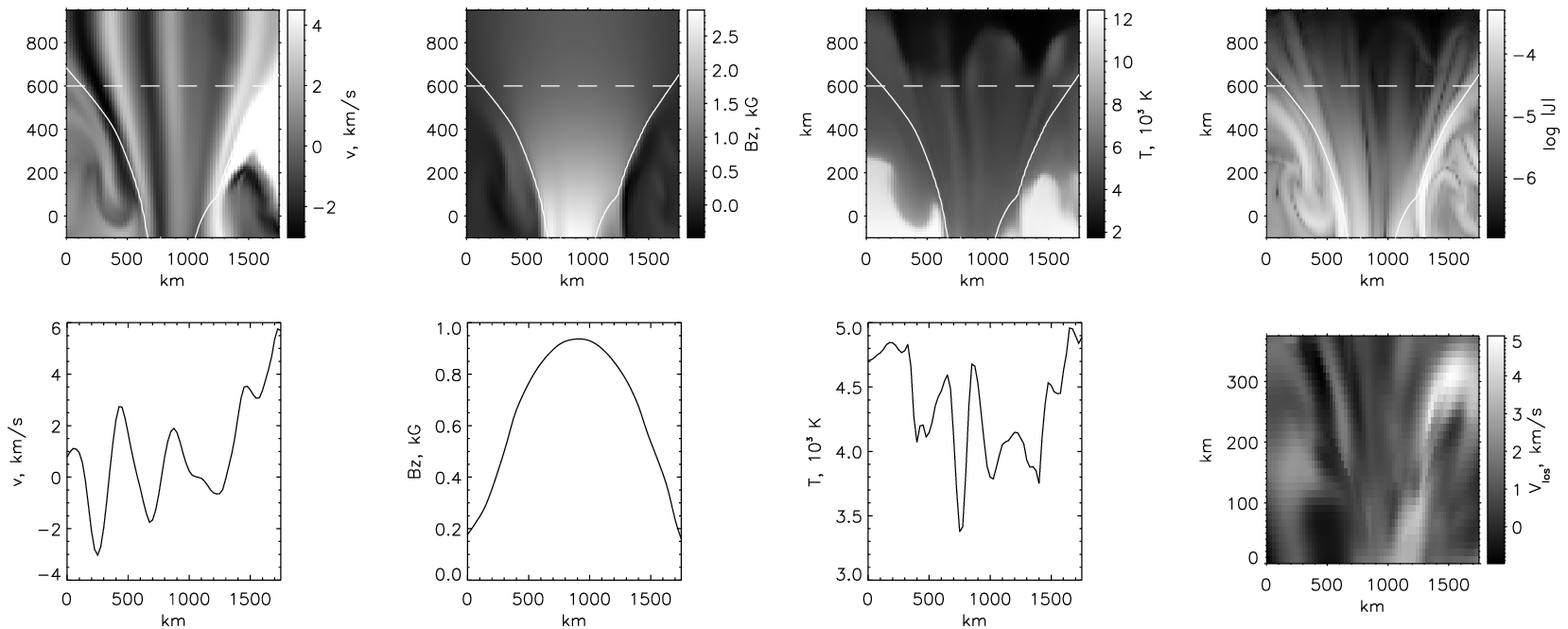} 
 \end{center}
\caption{Magneto-hydrodynamic parameters of a simulated photospheric "vortex" structure, marked by the box in Fig.~\ref{fig1}. Top row: the horizontal velocity (the component directed to 
the observer), the vertical component of magnetic field, the temperature, and the logarithm of the modulus of current density $\mathbf{J}=\nabla \times \mathbf{B}$ 
vertical cuts through the "vortex". Bottom row: the corresponding cuts of the aforementioned quantities, measured at $600~\mathrm{km}$ above the 
approximate level of the continuum formation, and the FeI 6302\AA~absorption line Doppler shift map for the region. Note that the Doppler shift map is rotated
counter-clockwise by $90^\mathrm{o}$ to align with other quantities in this figure. The contours in the top row bound the region with plasma $\beta<1$.}
\label{fig2}
\end{figure*}

An example of the calculated parameters of the simulated Stokes profiles at different inclinations is shown in Fig.~\ref{fig1}. The top row illustrates variation
of the emergent continuum intensity at $6303~\mathrm{\AA}$ with observational angle. As can be seen from a comparison with the second row of the figure, 
which is the line-of-sight component of magnetic field, the photospheric magnetic bright points \citep{shelyag2004} are clearly visible in the continuum at 
$0^\mathrm{o}$ inclination (left column of Fig.~\ref{fig1}, although with reduced contrast if compared to G-band observations. They gradually disappear towards the 
solar limb and are replaced by the hot and bright granular walls and faculae \citep{carlsson2004}. The continuum intensities are normalized by their 
mean value over all angles, thus the limb darkening is clearly visible.

In the third row of the figure, the Doppler shifts of FeI $6302.5~\mathrm{\AA}$ are shown. These are calculated as the shifts of their centres-of-gravity
(COG) from the reference zero-velocity wavelength $\lambda_0$, according to
\begin{equation}
\Delta \lambda = \lambda_{cog} - \lambda_0 = \frac{\int (I_c-I) \lambda d\lambda}{\int (I_c-I) d\lambda}-\lambda_0.
\label{linecog}
\end{equation}
As it can be seen in the plots, the structures in COG shift clearly resemble the photospheric convection velocity patterns. At $0^\mathrm{o}$, the granular
upflows are identified as negative, blue-shifted regions, while the intergranular downflows are positive and red-shifted. This situation changes at
non-zero inclinations. At $30^\mathrm{o}$ (middle column), the divergent granular flows show as blue-shifted at the sides of the granules facing 
the observer, and as red-shited regions at those facing away from the observer. 
At a $60^\mathrm{o}$ observational angle (right column), the corrugated 
optical surface of solar granulation hides the intergranular downflows, while divergent granular flows become significantly more pronounced. However, plenty of 
small-scale features, nearly co-located with the strong magnetic field regions (second row of the figure), emerging from the intergranular lanes 
and aligned with the direction of inclination, appear. Notably, no such structures in the continuum intensity images are found. These structures in 
the COG Doppler shift of FeI $6302.5~\mathrm{\AA}$ demonstrate observable presence of small-scale horizontal flows in the magnetised intergranular 
lanes of the simulation. One of these structures is marked by the box in Fig.~\ref{fig1} and used for further investigation.

The fourth row of Fig.~\ref{fig1} shows Stokes-$V$ area asymmetries, which are calculated by integrating Stokes-$V$ profiles over wavelength and 
normalising it by the total Stokes-$V$ areas, according to
\begin{equation}
A_V = \frac{\int V d\lambda}{\int |V| d\lambda}.
\label{Varea}
\end{equation}
As the figures show, at the solar disk centre (left column), the regions with the largest Stokes-$V$ asymmetry are the boundaries between the magnetised
intergranular lanes and the weakly-magnetised granules. This is due to the presence of strong gradients of the line-of-sight velocity and magnetic field components
in these regions \citep{illing1975, grossman1988, solanki1989, khomenko2005, shelyag2007}. The situation remains nearly unchanged for $30^\mathrm{o}$
inclination (middle row), while the regions with larger asymmetries are more pronounced and occupy larger areas, possibly due to the same mechanism of
Stokes-$V$ asymmetry generation. At $60^\mathrm{o}$, however, in a similar manner to the COG Doppler shift discussed above, together with spatially
large areas of enhanced asymmetry generated in granular regions, the small-scale elongated structures appear.

A similarity between the horizontal velocity structure in the simulations, FeI $6302.5~\mathrm{\AA}$ line Doppler shift and its Stokes-$V$ area asymmetry 
(Fig.~\ref{fig1}) can be demonstrated with Fig.~\ref{fig2}. In the top row of the figure, the horizontal velocity component (directed towards the observer), 
the vertical component of magnetic field, the temperature, and the modulus of current density $\mathbf{J}=\nabla \times \mathbf{B}$ are shown for vertical 
cuts through the magnetic field concentration marked by the box in Fig.~\ref{fig1}. The contours bound the regions with plasma $\beta<1$. 
In the bottom row of the figure, the horizontal velocity, the vertical magnetic field and the temperature are shown for a height of $600~\mathrm{km}$ above the 
approximate level of continuum formation in the magnetic flux tube, marked by 0 height in the top row. The bottom-right panel in the figure is the COG 
Doppler velocity map of the selected region.

The top-left plot of the figure shows presence of plasma, moving horizontally in opposite directions (towards and away from the observer),
within the magnetic field concentration \citep{shelyag2011a}, where the vertical magnetic field reaches $0.9~\mathrm{kG}$, as shown in the 
top and bottom plots in the second column of Fig.~\ref{fig2}. The plasma in the intergranular magnetic field concentration experiences some 
heating (as shown in third column of the figure) due to Ohmic dissipation of currents \citep{moll2012} in the magnetic field concentration (top-right plot).

By comparison of the Doppler velocity map (bottom-right panel) and the horizontal velocity field structure (top-left panel) in Fig.~\ref{fig2}, 
a conjecture can be made that the horizontal velocity motions in the photospheric intergranular 
magnetic field concentrations manifest themselves as the COG Doppler shifts of FeI $6302.5~\mathrm{\AA}$ absorption line and produce 
Stokes-$V$ asymmetry at large observational angles (see Fig.~\ref{fig1}). The asymmetry can be produced by a flow and magnetic field, simultaneously oscillating
in the region of the line formation. As suggested by \citet{shelyag2013}, the photospheric magnetic "vortices" are short-lived, oscillatory, 
transient events, which are manifestations of torsional Alfv\'en-type oscillations. 

\section{Degraded images}

\begin{figure}
  \includegraphics[width=9cm]{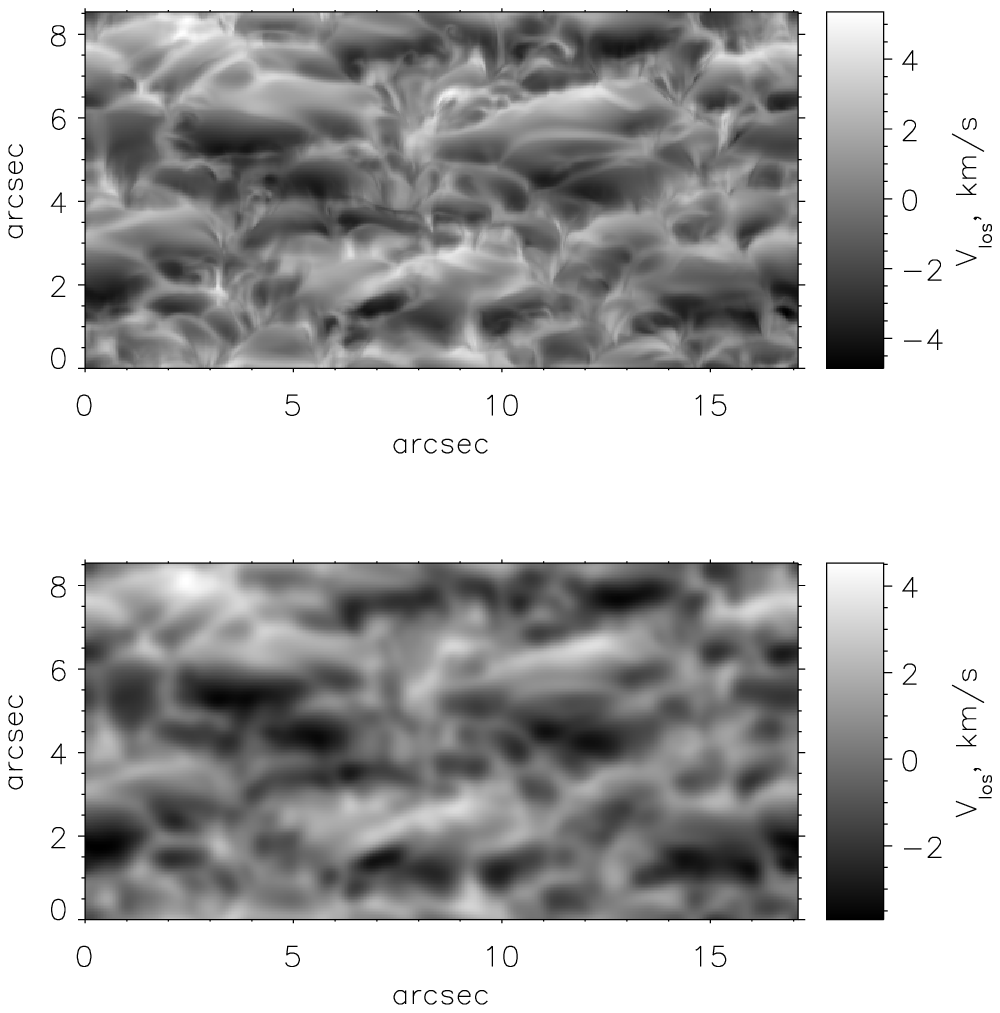} 
\caption{Original (top) and degraded (bottom) maps of line-of-sight velocities of} FeI $6302.5~\mathrm{\AA}$, computed for the simulated photosphere.
\label{fig3}
\end{figure}

Finally, we performed the simulated image degradation in a manner approximately representing Hinode Solar Optical Telescope (SOT) observations. 
The calculated Stokes-$I$ profiles
were convolved with Gaussian functions with $\sigma=0.2''$ for spatial resolution and with $\sigma=0.0215~\mathrm{\AA}$ for spectral resolution \citep{tsuneta1}. After
the degradation, the profiles were subjected to the same procedure of COG Doppler shift calculation. No instrumental noise has been added in this modelling. 
The original (non-degraded) and degraded images
are shown in upper and lower panels of Fig.~\ref{fig3}. As is evident from comparison of the degraded and original image, the small-scale 
structures in Doppler velocities completely disappear after the image degradation, while granular structure is clearly visible. This result suggests that higher-resolution observations 
are needed for the detection of photospheric Alfv\'en waves. The original resolution of the images was $25~\mathrm{km}$ ($\sim 0.04~\mathrm{arcsec}$),
which is the planned spatial resolution of the Advanced Technology Solar Telescope (ATST). Thus, future large-aperture instruments will be able to resolve this small-scale photospheric process.

However, it should be noted that the FeI $6302.5~\mathrm{\AA}$ line can be not the best choice for Alfv\'en wave detection. This line is formed in the low to 
mid photosphere, with the maximum of contribution function located at about $200~\mathrm{km}$ above the continuum formation layer
\citep{sanchez2000,khomenko2007,vigeesh2011}. Therefore, this line does not track the rich horizontal flow structure in the higher photosphere, as shown in Fig.~\ref{fig2}.
Including more realistic non-LTE effects in spectral line profile synthesis would lead to even deeper FeI line formation in the photosphere, as demonstrated by
\citet{schukina2001}. Nevertheless, it is suggested that torsional motions in the photospheric plasma, detected using local correlation tracking in observations, can be 
related to generation of torsional Alfv{\'e}n waves \citep{matsumoto1}.

\section{Conclusions}

In this paper, we performed spectro-polarimetric diagnostics of a solar photospheric magneto-convection model at three different observation angles.
The primary aim for this study was to detect possible observational signatures for horizontal, torsional, Alfv\'en-type motions in photospheric magnetic 
flux tubes. Using a photospheric absorption line of iron (FeI $6302.5~\mathrm{\AA}$), we found such signatures in the centre-of-gravity Doppler line shift,
as well as in the Stokes-$V$ area asymmetries. As it is shown, the perturbations, introduced in the Stokes profiles by torsional plasma motions in the low
plasma-$\beta$ intergranular
magnetic field concentrations, are observed only at high inclination angles. These features are found to be of very small spatial scales, less than the size of 
intergranular lanes. Aiming to confirm this finding at an observable resolution, we degraded the obtained images in order to mimic properties
of Solar Optical Telescope (SOT) onboard the Hinode satellite. As expected, the small-scale features disappeared after degradation. However, as the non-degraded
images suggest, it would be possible to detect the Alfv\'en wave signatures using an instrument with spatial resolution of about $0.04~\mathrm{arcsec}$.
Such spatial resolution is expected to be delivered by ATST, currently under construction. 

Another option would be a more appropriate 
choice of photospheric absorption line. FeI $6302.5~\mathrm{\AA}$ is formed deep in the solar photosphere, therefore it is unable to track upper-photospheric
regions, where larger-scale torsional motions in the magnetic flux tubes are demonstrated in the simulations, due to expansion of the magnetic flux tubes with
height. 

Finally, the result, presented in this paper, should be considered as another confirmation of the urgent need for observational instruments with larger 
apertures and higher resolutions for solar physics research.

\section{Acknowledgements}

This research was undertaken with the assistance of resources provided at the NCI National Facility systems at
the Australian National University, supported by Astronomy Australia Limited, and at the Multi-modal Australian 
ScienceS Imaging and Visualisation Environment (MASSIVE) (www.massive.org.au). The authors also gratefully thank Centre 
for Astrophysics \& Supercomputing of Swinburne University of Technology (Australia) for the computational resources 
provided. Dr Shelyag is the recipient of an Australian Research Council’s Future Fellowship (project number FT120100057).
The authors also thank the anonymous referee for a number of useful comments and suggestions which greatly improved the 
manuscript.

\end{document}